\documentclass[a4paper]{jpconf}
\usepackage{graphicx}
\usepackage{hyperref}

\newcommand{\nue}{\ensuremath{\nu_e}}
\newcommand{\nuebar}{\ensuremath{\bar{\nu}_e}}
\newcommand{\nux}{\ensuremath{\nu_x}}

\begin{document}
\title{Simulating fast time variations in the supernova neutrino flux in Hyper-Kamiokande}

\author{Jost Migenda for the Hyper-Kamiokande proto-collaboration}

\address{University of Sheffield, Department of Physics and Astronomy, Sheffield S3~7RH, UK}

\ead{jmigenda1@sheffield.ac.uk}

\begin{abstract}
Hyper-Kamiokande is a proposed next-generation water Cherenkov detector.
If a galactic supernova happens, it will deliver a high event rate ($\mathcal{O}(10^5)$ neutrino events in total) as well as event-by-event energy information. 
Recent supernova simulations exhibit the Standing Accretion Shock Instability (SASI) which causes oscillations in the number flux and mean energy of neutrinos.
The amplitude of these oscillations is energy-dependent, so the energy information available in Hyper-Kamiokande could be used to improve the detection prospects of these SASI oscillations.
To determine whether this can be achieved in the presence of detector effects like backgrounds and finite energy uncertainty, we have started work on a detailed simulation of Hyper-Kamiokande's response to a supernova neutrino burst.
\end{abstract}

\section{Introduction}
Hyper-Kamiokande (HK,~\cite{DiLodovico2016}) is a proposed next-generation water Cherenkov detector consisting of two cylindrical tanks with a fiducial volume of $187\,$kt each.
The first tank is expected to start data-taking in 2026, with the second tank following six years later.

HK has a broad physics programme covering many areas of particle and astroparticle physics, including oscillation of accelerator and atmospheric neutrinos, nucleon decay, solar neutrinos, supernova burst and relic neutrinos, indirect detection of dark matter through WIMP annihilation and neutrino geophysics.

For a galactic supernova at a fiducial distance of $10\,$kpc, HK will detect $\mathcal{O}(10^5)$ neutrinos within 10\,s.
This high event rate enables HK to resolve fast time variations of the event rate, which could give us information on properties of the progenitor (like its rotation) or on details of the supernova explosion mechanism which are currently still unclear.

In section~\ref{sec:sasi} of this paper, we will describe a feature of supernova explosions that causes such fast time variations and we introduce a method of using this energy information.
To determine whether this method can be successful in the presence of detector effects, we have started work on a detailed detector simulation, which is described in section~\ref{sec:simulation}.

\section{The standing accretion shock instability}\label{sec:sasi}
\subsection{Previous work}
To date, only two dozen supernova burst neutrinos have been detected, all from SN1987a~\cite{SN1987a}, and progress in our understanding of the supernova explosion mechanism has come mostly from computer simulations.
In 2003, Blondin and others~\cite{Blondin2003} reported that small perturbations of a spherical shock front can be self-sustaining and lead to rapidly growing turbulence behind the shock.
This feature, which has been called standing accretion shock instability (SASI), has since played a crucial role in facilitating explosions in some two- and three-dimensional supernova simulations, but remained subdominant in others, leaving the role it plays in realistic supernova explosions the subject of debate~\cite{Hanke2013,Burrows2012}.
Observing it would allow us to discriminate between different supernova models and improve our understanding of the explosion mechanism.

SASI is characterized by oscillations of both the number flux and the mean energy of supernova burst neutrinos with a frequency of $\mathcal{O}(100)\,$Hz over a period of a few hundred ms.
Previous work found that for an ideal scenario, these SASI oscillations of the number flux are clearly detectable in both IceCube and HK~\cite{Lund2010,Tamborra2013,Tamborra2014}.
However, the amplitude of these oscillations is direction-dependent and may be much smaller if the detectors on Earth are at a large angle to the SASI plane.
To improve our chances of detecting SASI in such a scenario, we can make use of the additional event-by-event energy information that is available in HK but not in IceCube.

\subsection{Energy-dependence of SASI oscillations}
We use data from a 
simulation by the Garching group~\cite{Hanke2013,Tamborra2013} of a $27\,M_\odot$ progenitor~\cite{Woosley2002}.
The preprocessed data set used here (see appendix~A of reference~\cite{Tamborra2014}) contains the mean energy, mean squared energy and luminosity of \nue, \nuebar\ and \nux\ (all other flavors) during the first 550\,ms after the core-bounce for the observer direction with maximal SASI amplitude.


\begin{figure}[h]
\begin{minipage}{17pc}
\includegraphics[width=19pc]{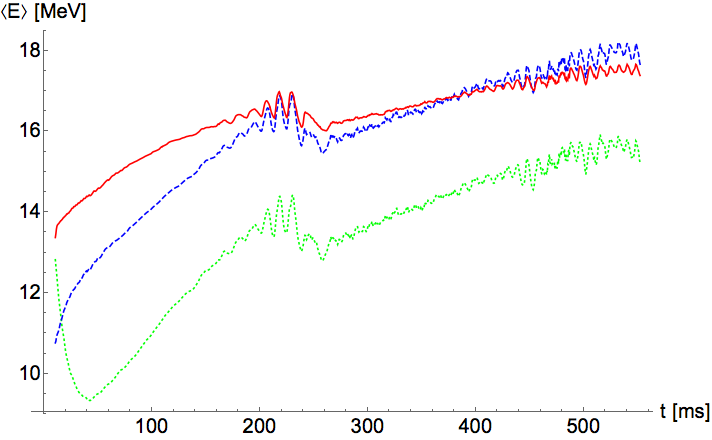}
\end{minipage}\hspace{2pc}
\begin{minipage}{17pc}
\includegraphics[width=19pc]{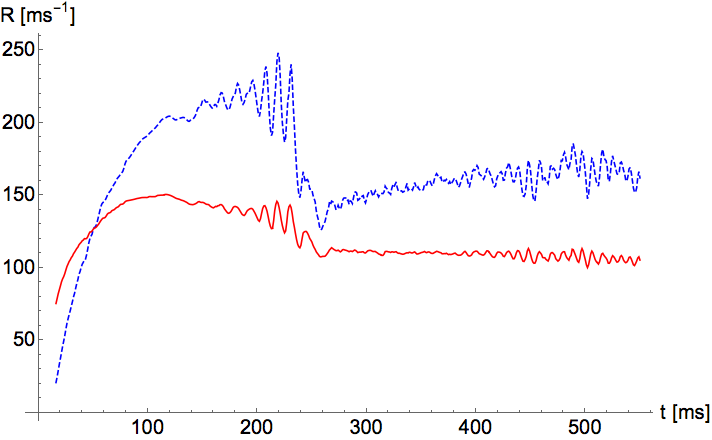}
\end{minipage} 
\caption{Left: Mean energy of \nue\ (green, dotted), \nuebar\ (blue, dashed) and \nux\ (red, solid). Right: Event rate in HK without oscillations (blue, dashed) and with full flavor swap between \nuebar\ and \nux\ (red, solid). SASI oscillations are visible at around 220\,ms and 500\,ms.}
\label{fig:sasi-oscillations}
\end{figure}

Figure~\ref{fig:sasi-oscillations} displays the event rate expected in HK and the mean neutrino energy.
The SASI oscillations of both are in phase.
Therefore, when the total event rate is high, the mean energy is also high, leading to a very high number of high-energy events; while when the total event rate is low, the mean energy is also low, leading to a very low number of high-energy events.

Accordingly, the amplitude of the SASI oscillations of the event rate will be larger in the highest-energy bin than in the total signal.
On the other hand, that high-energy bin contains fewer events, leading to a higher relative shot noise.
Overall, considering the high-energy bin instead of the total signal is advantageous if the oscillations of the mean energy detected in HK are slightly higher than those of \nuebar\ in this supernova simulation~\cite{Migenda2015}.

\section{Detector simulation}\label{sec:simulation}
After showing that it is in principle possible to use the event-by-event energy information available in HK, we need to determine whether this can realistically be achieved in the presence of detector effects like backgrounds and finite energy uncertainty.
To investigate this, we have started work on a detailed simulation of HK's response to a supernova neutrino burst.

We extract the neutrino flux from the data set described in section~\ref{sec:sasi} and multiply it with the energy-dependent inverse beta decay (IBD) cross-section~\cite{Strumia2003} to get the number of IBD events.
Since IBD is the dominant process for supernova neutrinos in HK, making up about 90\,\% of the events, we scale this by a factor of $0.9^{-1}$ to get an approximation for the total number of events.

\begin{figure}[h]
\includegraphics[width=17.4pc]{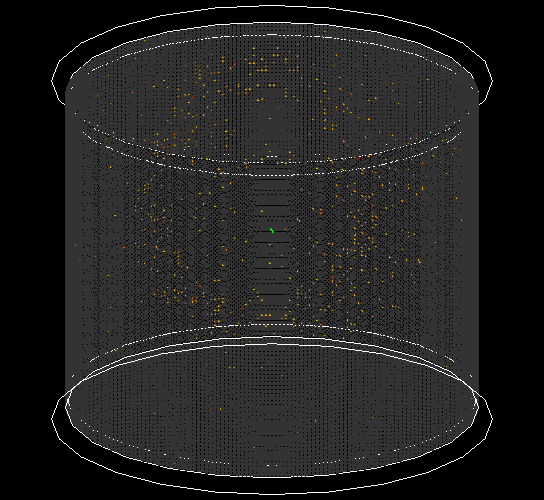}\hspace{2pc}%
\begin{minipage}[b]{14pc}\caption{\label{fig:wcsim}Signal of a 50\,MeV positron from an inverse beta decay event in HK, simulated by WCSim. The green dot at the centre of the detector marks the location of the event, while the orange and red dots signify PMT hits. A faint Cherenkov ring is visible.}
\end{minipage}
\end{figure}

We distribute the resulting positrons randomly within the detector volume and simulate the detector response with WCSim~\cite{WCSim}, a Geant4-based application for simulating water Cherenkov detectors (see figure~\ref{fig:wcsim}).
We also simulate backgrounds, most importantly dark noise events from the PMTs.
For reconstruction we use BONSAI, a software fitter that was originally developed for Super-Kamiokande~\cite{Smy2007}. 

\section{Summary}\label{sec:summary}
In a galactic supernova, SASI can lead to oscillations of the neutrino number flux that are detectable in HK.
The amplitude of these oscillations is larger for higher neutrino energies and under some conditions, it is theoretically advantageous to consider just a subset of the signal.
To test whether this holds up in the presence of detector effects, we are working on a detailed simulation of HK's response to a supernova neutrino burst and expect to present first results soon.

\end{document}